\documentclass[aps,prl,reprint,nofootinbib,showpacs,superscriptaddress,twocolumn,longbibliography]{revtex4-1}

\usepackage[utf8]{inputenc}
\usepackage{braket}
\usepackage{algorithm}
\usepackage{xcolor}
\usepackage{mathtools}
\usepackage{graphicx}
\usepackage{breqn}
\usepackage{tabularx}
\usepackage[noend]{algpseudocode}
\usepackage{amsmath}
\DeclarePairedDelimiter\floor{\lfloor}{\rfloor}
\newtheorem{theorem}{Theorem}
\DeclareMathOperator{\IQFT}{IQFT}
\newcommand{\tens}[1]{%
  \mathbin{\mathop{\otimes}\limits_{#1}}%
}

\makeatletter
\newcommand{\multiline}[1]{%
  \begin{tabularx}{\dimexpr\linewidth-\ALG@thistlm}[t]{@{}X@{}}
    #1
  \end{tabularx}
}
\let\cat@comma@active\@empty
\makeatother
\usepackage[english]{babel}

\begin{document}

\title{Quantum Medical Imaging Algorithms}

\author{Bobak Toussi Kiani}
\author{Agnes Villanyi}
\author{Seth Lloyd}
\affiliation{Massachusetts Institute of Technology, 77 Massachusetts Avenue, Cambridge, MA 02139, USA}

\begin{abstract}
A central task in medical imaging is the reconstruction of an image or function from data collected by medical devices (\textit{e.g.,} CT, MRI, and PET scanners). We provide quantum algorithms for image reconstruction with exponential speedup over classical counterparts when data is input as a quantum state. Since outputs of our algorithms are stored in quantum states, individual pixels of reconstructed images may not be efficiently accessed classically; instead, we discuss various methods to extract information from outputs using a variety of quantum post-processing algorithms.
\end{abstract}

\maketitle

\section{Introduction}

Image reconstruction algorithms are used in many fields to construct visual representations from input data collected by a device. In medical imaging, these devices include Magnetic Resonance Imaging (MRI), Computed Tomography (CT), and Positron Emission Tomography (PET) scanners \cite{hsiehComputed2009, cormackRepresentation1963, aundalRadon1996, liangPrinciples1999}. Image reconstruction algorithms often take advantage of relations between a target image or function and its representation in the frequency (Fourier) domain. Based on the method by which data is collected (and in which domain), image reconstruction algorithms can be divided into two categories. The first category pertinent to MRI scanners are algorithms to reconstruct images from input data collected in the frequency domain called k-space in MRI \cite{liangPrinciples1999}. The prototypical reconstruction algorithm in this case is a simple inverse Fourier transform. The second category pertinent to CT and PET scanners are algorithms to reconstruct images from a set of projections or line integrals over a function. Image reconstruction in this case can be mathematically formulated as an inverse Radon transform where a wide range of algorithms can be used to perform reconstruction. Perhaps the most fundamental among these algorithms are implementations of the Fourier Slice Theorem \cite{hsiehComputed2009,aundalRadon1996}.

In this study, we provide quantum image reconstruction algorithms for both categories described above. A wide variety of quantum data processing algorithms take as input a quantum state representing the data and process it using a quantum computer \cite{harrowQuantum2009,le2010fast,wiebeQuantum2012,dallaire-demersQuantum2018,caraimanQuantum2013,biamonteQuantum2017,lloydQuantum2014,sasakiQuantum2001,berry2014high}. The quantum medical imaging algorithms proposed here take, as input, a quantum state representing the data outputted from a medical imaging device and reconstruct "quantum images" -- quantum states in a superposition of pixel values. Since the algorithms' outputs are quantum states, reading out all the pixels in a quantum image in general may not be efficient. However, outputs of quantum reconstruction algorithms can be subsequently post-processed using other quantum algorithms and techniques. For example, a large body of literature exists providing efficient algorithms for image processing using quantum computers \cite{yanQuantum2017,hoyerEfficient1997,caraiman2012image,beach2003quantum,le2010fast,jiang2015quantum}. In addition, many quantum machine learning and data processing algorithms that take quantum images as input provide exponential speedups over their classical counterparts \cite{hoyerEfficient1997,klappenecker2001discrete,lloydQuantum2014,biamonteQuantum2017,liuQuantum2018,caraiman2012image,lloydQuantum2018}.

Implementing image reconstruction algorithms in quantum computers offers two unique advantages over the classical setting. First, quantum image reconstruction algorithms can be run more efficiently, in many cases requiring only poly-logarithmic time with respect to the size of an image (or number of pixels). Second, quantum algorithms for image reconstruction perform operations on a quantum wavefunction (as opposed to classically sampled data) which opens the possibility of collecting input data in a quantum mechanical manner using potentially less time or smaller doses of radiation. 

This study is organized as follows. First, we show how the prototypical algorithm for image reconstruction in the case of MRI scans gives rise to a simple and efficient quantum algorithm. Then, we provide a short theoretical overview of the Radon transform for the case of CT and PET scans. In this setting, algorithms for reconstruction via an implementation of the Fourier Slice Theorem are detailed and contrasted both for classical and quantum computation. Finally, to outline the different means of obtaining information from outputted quantum states, we show how to apply different methods for post-processing quantum images to extract useful information from the output quantum states.

\section{Magnetic Resonance Imaging}

Algorithms for image reconstruction from MRI scanners reconstruct the aggregate density of nuclear spins in a subject being scanned. Interactions between nuclear spins and externally applied magnetic fields cause a bulk precession of nuclear spins. The signals generated by the spins are obtained by recording the voltages of receiving coils inductively coupled to the magnetization. In its standard form, MRI data is collected in the Fourier spectrum of the density function that is being reconstructed (commonly termed k-space). The signal $s$ received by an MRI scanner can be related to the density of nuclear spins  $\rho$ using a Fourier transform \cite{liangPrinciples1999, haackeMagnetic1999}:

\begin{equation}
    s(k_x,k_y) = \int \int \rho (x,y) e^{-i2\pi (k_x x + k_y y)} \,dx \,dy
    \label{eq:MRIContinuous}
\end{equation}{}

where $x$ and $y$ represent two dimensional spatial variables and $k_x$ and $k_y$ represent the corresponding frequency variables for those dimensions. The relation above can also be extended to reconstruction of one dimensional or three dimensional functions \cite{liangPrinciples1999, haackeMagnetic1999}.

Data at different frequencies in k-space is collected by applying linear gradients to the magnetic field along targeted directions and taking measurements at different times. The relation between the frequencies $(k_x, k_y)$ and the spatial gradients of the magnetic field $(G_x, G_y)$ is time dependent and given by \cite{liangPrinciples1999}:

\begin{equation}
\label{eq:gradient}
\begin{split}
    & k_x(t) =  \frac{\gamma}{2\pi} \int_0^{t} G_x(t')\,dt' \\
    & k_y(t) =  \frac{\gamma}{2\pi} \int_0^{t} G_y(t')\,dt'
\end{split}
\end{equation}
where $\gamma$ is the gyromagnetic ratio.

In the prototypical arrangement, MRI data is sampled uniformly in the frequency domain. In this case, reconstruction of the density of nuclear spins $\rho$ is a simple inverse fast Fourier transform (see equation \ref{eq:MRIContinuous}) which would require $O(N^2 \log N)$ time to reconstruct an $N \times N$ image classically.

\subsection{A Quantum Algorithm for Magnetic Resonance Imaging}

Since MRI data is collected in the frequency domain (k-space), the quantum algorithm for MRI image reconstruction is particularly simple. Here, we assume the input to the quantum algorithm is a quantum state $s(k_x,k_y)\ket{k_x}\ket{k_y}$ containing signal amplitudes of the MRI data collected in the frequency domain. The frequency space is indexed by ket vectors for each dimension $\ket{k_x}$ and $\ket{k_y}$. As evident from equation \ref{eq:MRIContinuous}, when the k-space is sampled uniformly in the two frequency dimensions, the algorithm to recover $\rho (x,y) \ket{x} \ket{y}$ is a simple 2-D inverse Fourier transform of the data in k-space.

\begin{equation}
    \rho (x,y) \ket{x} \ket{y} = \IQFT \, s(k_x,k_y)\ket{k_x}\ket{k_y}
\end{equation}
where $\IQFT$ is the inverse quantum Fourier transform in 2 dimensions and $\ket{x}$ and $\ket{y}$ index the spatial dimensions of the image.

The total run-time for this reconstruction algorithm on a quantum computer corresponds to the time complexity of a quantum Fourier transform: $O(\log^2 N)$ to reconstruct an $N \times N$ image of the density function.

\section{Radon Transform and Computed Tomography}

Image reconstruction from Computed Tomography (CT) and Positron Emission Tomography (PET) scans combine flux measurements over a set of angles to reconstruct cross-sections of the subject being scanned \cite{hsiehComputed2009, cormackRepresentation1963, aundalRadon1996}. We propose a quantum algorithm to efficiently reconstruct a two dimensional function from sets of parallel line integrals over that function. In the case of CT scans, this function characterizes the linear attenuation coefficients of the object being scanned indicating how much light passes through the object \cite{hsiehComputed2009}. In the case of PET scans, this function characterizes the radioactive tracer (radionuclide) concentrations within a biological specimen \cite{aundalRadon1996, cherryPet2006}.

Mathematically, the Radon Transform returns line integrals over a function at specified angles. In the case of tomographic image reconstruction, if $F(x,y)$ specifies the linear attenuation of an object in two spatial dimensions, then the Radon Transform returns the line integral of those coefficients at specified angles and linear offsets. 

In practice, a Radon transform mathematically formulates the projection data that is collected by a device (e.g. in CT scans). A reconstruction algorithm takes as input the data from a Radon transform applied on a function $F(x,y)$ and reconstructs another function $G(x,y)$ that is close to the original function $F(x,y)$. Many different algorithms exist to perform this image reconstruction \cite{hsiehComputed2009, aundalRadon1996, pipatsrisawatPerformance2005, pressDiscrete2006}. The specific algorithm we consider performs reconstruction via an implementation of the Fourier Slice Theorem. Though this implementation is not commonly used in the medical imaging community since it requires interpolation in the frequency domain, it nonetheless serves as a theoretical starting point for image reconstruction algorithms in general \cite{hsiehComputed2009,aundalRadon1996}. More commonly used algorithms for image reconstruction include those that perform interpolation in the spatial domain (filtered back-projection) \cite{pipatsrisawatPerformance2005}, invert a discrete version of the radon transform \cite{pressDiscrete2006, beylkinDiscrete1987}, or use iterative procedures to perform image reconstruction \cite{aundalRadon1996, hsiehComputed2009}. It is an open question whether quantum computers offer an exponential speedup for these other algorithms. A brief description of these other algorithms is included in the supplemental materials.

A radon transform $\mathcal{R}$, which takes as input a function $F(x,y)$ and returns the line integral of that function over a specified line, can be written in various different equivalent forms. One common form is below:

\begin{dmath}
    \mathcal{R}F(x,y) = f(\rho,\theta) = \int_{-\infty}^{\infty} \int_{-\infty}^{\infty} F(x,y) \delta(x\cos{\theta}+y\sin{\theta}-\rho) \,dx\,dy
    \label{eq:radonTransform}
\end{dmath}
Equation \ref{eq:radonTransform} can be interpreted as the integral of the function $F(x,y)$ along the line $x\cos{\theta}+y\sin{\theta}=\rho$ \cite{aundalRadon1996}. Throughout this study, we use $G(x,y)$ to indicate the reconstructed function which ideally approximates the true function $F(x,y)$. $f(\rho,\theta)$ indicates the result of the Radon transform applied to $F(x,y)$. 

Our quantum algorithm reconstructs $G(x,y) \approx F(x,y)$ from $f(\rho,\theta)$ via an implementation of the Fourier slice theorem (also known as the projection-slice theorem) \cite{bracewellNumerical1990}. For simplicity, we consider the case of image reconstruction in two spatial dimensions; an extension to three dimensions is provided in the supplementary materials.

\begin{theorem}[Fourier slice theorem \cite{hsiehComputed2009,aundalRadon1996}]
\label{slice}
Let $\hat{F}(k_x,k_y)$ be the two dimensional Fourier transform of  $F(x,y)$ and $\hat{f}(k_\rho,\theta)$ be the one dimensional Fourier transform of $f(\rho,\theta)$ over the $\rho$ dimension. The values of $\hat{f}(k_\rho,\theta)$ are equal to the values of $\hat{F}(k_x,k_y)$ on the slice $(k_x = k_\rho \cos \theta,k_y = k_\rho \sin \theta)$ passing through the origin at the same angle $\theta$. 
\begin{equation}
    \hat{F}(k_x,k_y) = \int_{-\infty}^{\infty} \int_{-\infty}^{\infty} F(x,y) e^{-i2\pi(k_x x+k_y y)} \,dx\,dy 
\end{equation}
\begin{equation}
    \hat{F}(k_\rho \cos \theta,k_\rho \sin \theta) = \int_{-\infty}^{\infty} f(\rho, \theta) e^{-i2\pi \rho k_\rho}\,d\rho
\end{equation}
\end{theorem}

\begin{figure}[ht]
    \centering
    \includegraphics[scale = 0.2]{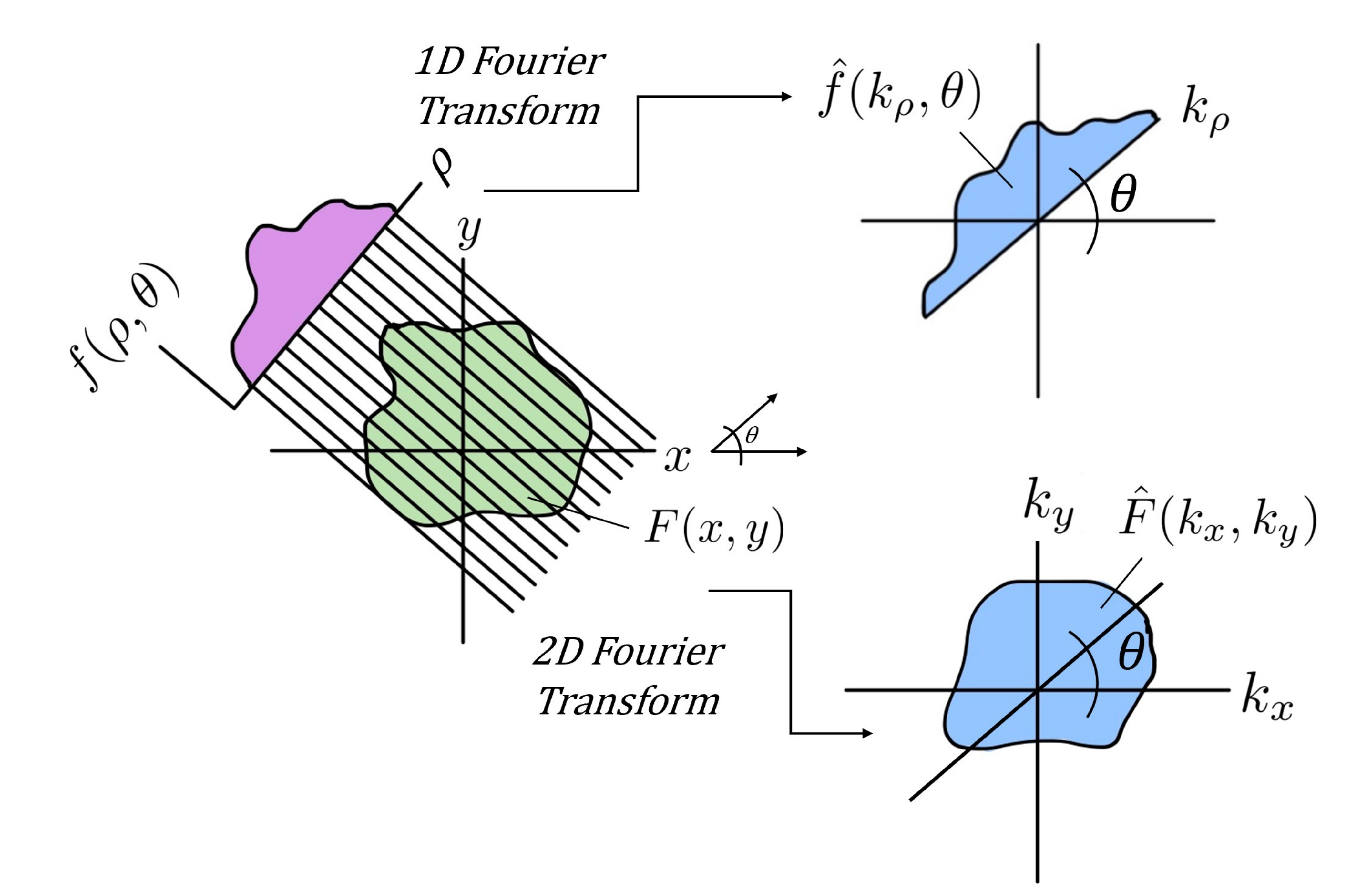}
    \caption{Schematic for Fourier slice theorem: the 1D Fourier transform of a projection at angle $\theta$ is equivalent to a slice of the original function's 2D Fourier transform at angle $\theta$.}
    
    \label{fig:fourierSlice}
\end{figure}

A visual interpretation of the Fourier slice theorem is shown in figure \ref{fig:fourierSlice} detailing the connection between the function in the spatial and frequency domains. When given discretized data, the goal is to determine the values of the 2D spectrum $\hat{F}(k_x,k_y)$ at discrete values of $k_x$ and $k_y$ and then invert the 2D spectrum to obtain a reconstructed image $G(x,y) \approx F(x,y)$. Based on the Fourier slice theorem, discrete values of $\hat{F}(k_x,k_y)$ at coordinates $k_x$ and $k_y$ can be calculated using interpolation from values of $\hat{f}(k_\rho, \theta)$ at slices that fall close to those coordinates. Algorithm \ref{alg:classical} details the steps in image reconstruction using the Fourier slice theorem when given discrete data.

\begin{algorithm}[H]
\caption{(Classical) Image Reconstruction via Fourier Slice Theorem}\label{alg:classical}
\textbf{Input: }Set of projections: $f(\rho,\theta)$\\
\textbf{Result: }Reconstructed image: $G(x,y)$
\begin{algorithmic}[1]
\State 1D FFT on $\rho$ dimension \Comment{returns $\hat{f}(k_\rho,\theta)$}
\label{alg:classical1dfft}
\State interpolate values of $\hat{F}(k_x,k_y)$ from slices ($\hat{f}(k_\rho,\theta)$ is in polar form)
\label{alg:classicalInterpolation}
\Comment{returns $\hat{F}(k_x,k_y)$}
\State \textbf{return} inverse 2D FFT on $\hat{F}(k_x,k_y)$ as reconstructed image in spatial coordinates $G(x,y)$
\label{alg:classical2dfft}
\end{algorithmic}
\end{algorithm}

Various different methods of interpolation are available for step \ref{alg:classicalInterpolation} of algorithm \ref{alg:classical}. Some of these are detailed in the supplemental materials.

To reconstruct an $N \times N$ image given sets of parallel projections at $N$ discrete angles, the classical image reconstruction algorithm based on the Fourier slice theorem has $O(N^2 \log N)$ cost. This cost is dominated by the Fourier transform steps. Other commonly used methods of classical image reconstruction also require at least $O(N^2 \log N)$ time \cite{pipatsrisawatPerformance2005, aundalRadon1996, hsiehComputed2009}.

\subsection{Quantum Implementation of Fourier Slice Theorem}

Below, we present a quantum algorithm for image reconstruction of an $N \times N$ image that runs in time $O(s^2 \log{N} + \log ^2 N )$ in cases where the frequency data is well conditioned as described in the supplementary materials (\textit{i.e.,} Fourier transform of projections is not dominated by the low frequencies). Here, $s$ is a constant that indicates the number of points used to interpolate data from polar to Cartesian coordinates (does not depend on $N$). This is an exponential improvement over the classical runtime of $O(N^2 \log N )$.

The quantum algorithm follows almost directly from the classical image reconstruction algorithm. We assume that the input data $f(\rho, \theta)$ is provided as a quantum state in two registers: $f(\rho,\theta)\ket{\rho}\ket{\theta}$. The $\ket{\rho}$ and $\ket{\theta}$ registers indicate the discrete coordinates of the $\rho$ and $\theta$ dimensions respectively equally spaced in $\rho$ and $\theta$, whereas the normalized value of $f(\rho,\theta)$ is encoded as the amplitude of the quantum state. 

The quantum algorithm is detailed in algorithm \ref{alg:IQR}. The first step of the algorithm is a Quantum Fourier Transform (QFT) on the $\ket{\rho}$ register. Next, we linearly interpolate from polar coordinates to Cartesian coordinates using a sparse linear interpolation Hamiltonian matrix. Finally, to reconstruct the image, we perform a 2D inverse QFT on the $\ket{x}$ and $\ket{y}$ registers.

\begin{algorithm}[H]
\caption{(Quantum) Image Reconstruction via Fourier Slice Theorem}
\label{alg:IQR}
\textbf{Input: }Set of projections: $f(\rho,\theta)\ket{\rho}\ket{\theta}$ \\
\textbf{Result: }Reconstructed image: $G(x,y) \ket{x} \ket{y}$
\begin{algorithmic}[1]
\State 1D QFT on $\ket{\rho}$ register \Comment{$\hat{f}(k_\rho,\theta)\ket{k_\rho}\ket{\theta}$}
\State append ancillary qubit 
\Comment{$\hat{f}(k_\rho,\theta)\ket{1}\ket{k_\rho}\ket{\theta}$}
\While{measurement ancillary qubit $=\ket{1}$}
\State \multiline{%
apply sparse interpolation Hamiltonian: $e^{-iA^* t}$}
\Comment{\textit{see description below}}
\label{alg:interp1}
\State \textit{optional: } perform amplitude amplification on ancillary qubit
\State measure ancillary qubit
\Comment{\textit{see description below}}
\label{alg:interp2}
\EndWhile
\State \textbf{return} inverse 2D QFT of interpolated Fourier matrix $\hat{F}(k_x,k_y) \ket{k_x}\ket{k_y}$ \Comment{$G(x,y)\ket{x}\ket{y}$}
\end{algorithmic}
\end{algorithm}

Using efficient algorithms for sparse Hamiltonian simulation \cite{berryEfficient2007}, applying the $s$-sparse linear interpolation Hamiltonian (step \ref{alg:interp1}) can be performed in $O(s^2 \log N)$ time. The quantum Fourier transforms take time $O(\log ^2 N)$. Thus, the runtime of the quantum image reconstruction algorithm is $O(s^2 \log{N} + \log ^2 N )$.

\subsection{Quantum Input and Output States}
Note that for our quantum implementation of the Fourier slice theorem, projections are assumed to be collected along parallel lines as shown in figure \ref{fig:fourierSlice} \cite{hsiehComputed2009, aundalRadon1996}. The input to our quantum image reconstruction algorithm is a quantum state in two registers $f(\rho,\theta)\ket{\rho}\ket{\theta}$. The $\ket{\rho}$ register indexes the offsets of the parallel line projections and the $\ket{\theta}$ register indexes the angles at which projections are taken.

The output of the image reconstruction algorithm is a discrete 2-dimensional array stored in a quantum state  that can be interpreted as a "quantum image". Various different methods exist to cast images into quantum states \cite{yanSurvey2016, caraimanQuantum2013, mastrianiQuantum2016}. In our quantum image reconstruction algorithm, the output reconstructed image ($G(x,y)\ket{x}\ket{y}$) has two registers indexing the discrete $x$ and $y$ spatial coordinates. This reconstructed image can be construed as a grayscale image, where the magnitude $G(x,y)$ at a specified set of coordinates indicates the pixel intensity of the reconstructed image. 

\subsection{Polar to Cartesian Interpolation (steps \ref{alg:interp1}-\ref{alg:interp2})}

Prior to interpolation, the data is in polar coordinates $\hat{f}(k_\rho,\theta)\ket{k_\rho}\ket{\theta}$, and our aim is to convert this data to Cartesian coordinates $\hat{F}(k_x,k_y) \ket{k_x}\ket{k_y}$. A sparse linear interpolation matrix $A$ is formed to perform this conversion, and we append an ancilla qubit in the $\ket{1}$ state to allow for post-selection ensuring that the interpolation is successful. For each discrete set of frequencies $k_x$ and $k_y$ in Cartesian coordinates, linear coefficients are stored in the matrix $A$ which uses up to $s$ entries of $\hat{f}(k_\rho,\theta)\ket{k_\rho}\ket{\theta}$ to calculate $\hat{F}(k_x,k_y)$ at any given $k_x$ and $k_y$ coordinates. In other words, the linear interpolation matrix mapping polar to Cartesian coordinates has at most $s$ entries per row. Different methods of choosing the $s$ entries and their values include nearest neighbor ($s=1$), simplex interpolation ($s=3$), and bilinear interpolation ($s=4$). These are further detailed in the supplementary materials. 

We implement sparse matrix multiplication on a quantum computer using sparse Hamiltonian simulation, which runs in time $O(s^2 \log N)$. In general, the matrix $A$ is not Hermitian; thus, we define the matrix $A^*$:
\begin{equation}
    A^* = \sigma^- \tens{} A + \sigma^+ \tens{} A^\dagger = 
    \begin{pmatrix}
    0 & A\\
    A^\dagger & 0
    \end{pmatrix}
\end{equation}

After forming $A^*$, we aim to perform the matrix multiplication below:
\begin{dmath}
    A^* \ket{1}\hat{f}(k_\rho,\theta)\ket{k_\rho}\ket{\theta} = 
    \begin{pmatrix}
    0 & A\\
    A^\dagger & 0
    \end{pmatrix}
    \begin{pmatrix}
    0\\
    \hat{f}(k_\rho,\theta)
    \end{pmatrix}
    =
    \begin{pmatrix}
    \hat{F}(k_x,k_y) \\
    0
    \end{pmatrix}
\end{dmath}

We cannot implement the above directly; however, using sparse Hamiltonian simulation techniques, we can apply the matrix $A^*$ as a Hamiltonian and obtain $\hat{F}(k_x,k_y)$ up to a given error $\epsilon$. Specifically, we apply the following Hamiltonian operator:

\begin{dmath}
    e^{-iA^* t} \ket{1}\hat{f}(k_\rho,\theta)\ket{k_\rho}\ket{\theta} = (I+O(t^2)A^{*\dagger}A^*)\ket{1}\hat{f}(k_\rho,\theta)\ket{k_\rho}\ket{\theta} + i (t+O(t^3)A^*A^{*\dagger})\ket{0}\hat{F}(k_x,k_y) \ket{k_x}\ket{k_y} 
    \approx \ket{1}\hat{f}(k_\rho,\theta)\ket{k_\rho}\ket{\theta} + it \ket{0}\hat{F}(k_x,k_y) \ket{k_x}\ket{k_y}
\end{dmath}

The time $t$ is chosen to ensure that the $O(t^3)$ error term is smaller than $\epsilon$; see supplementary materials for proof that $t$ does not depend on $N$. Finally, the ancilla qubit is measured and the process is repeated until the measurement of the ancilla qubit is in the $\ket{0}$ state. Assuming $\epsilon$ is small, the probability of successfully measuring the ancilla qubit in the state $\ket{0}$ is equal to $p_0 = t^2 \bra{k_y}\bra{k_x}\hat{F}(k_x,k_y)^\dagger \hat{F}(k_x,k_y) \ket{k_x}\ket{k_y}$. Optionally, one can perform amplitude amplification on the ancilla qubit to improve the probability of measuring $\ket{0}$. Using amplitude amplification, the process requires on average $O(\frac{1}{\sqrt{p_0}})$ iterations to perform matrix interpolation successfully \cite{brassardQuantum2000}. In the supplemental materials, we show that the interpolation can be efficiently performed in cases where the frequency data is not dominated by low frequencies.

\section{Post-processing of Quantum Reconstructed Images}

Outputs of the quantum image reconstruction algorithms proposed here are quantum states in a superposition of pixel locations. Directly reading the pixel values of a $N \times N$ reconstructed quantum image requires time $O(N^2)$. Applying quantum post processing algorithms to the quantum image allows us to obtain useful information in time $O(\textrm{poly}(\log{N}))$. We note that in cases where images can be compressed via an efficient transformation onto a given basis, even reading out pixel values can be efficient. For example, suppose that the image is highly compressible under the discrete cosine transform which forms the basis for JPEG compression \cite{wallace1992jpeg}. Then, applying the quantum version of the discrete cosine transform and measuring the components of the transformed quantum state will allow us to read out the compressed image \cite{pang2006quantum,klappenecker2001discrete}.

When one is interested in processing the quantum image to extract key information, images stored in the quantum states can be passed into to a host of other quantum algorithms for further analysis or processing. For example, a large array of quantum machine learning algorithms exist that may prove useful in analyzing reconstructed images \cite{biamonteQuantum2017}. Among the most promising of these quantum machine learning algorithms are those for neural networks (including convolutional neural networks) \cite{congQuantum2019,killoranContinuous-variable2019}, principal component analysis \cite{lloydQuantum2014}, generative adversarial networks \cite{dallaire-demersQuantum2018,lloydQuantum2018}, and anomaly detection \cite{liuQuantum2018}.

Images stored in a quantum state also offer the possibility for more efficient post-processing compared to classical computation. Many common image processing techniques are exponentially faster on a quantum computer; these include the Fourier transform and certain wavelet transforms such as the Haar transform and Daubechies’ $D^4$ transform \cite{hoyerEfficient1997,nielsenQuantum2011}.  In addition, algorithms for template matching have been proposed that may offer exponential speedup on a quantum computer \cite{sasakiQuantum2001,curtis2004towards}. These algorithms can be used individually or in combination with the machine learning algorithms discussed in the prior paragraph.

\section{Conclusion}

Our results provide efficient quantum algorithms for medical image reconstruction on quantum computers. If input data is provided as a quantum state, quantum algorithms can yield exponential speed-ups over their classical counterparts. Quantum algorithms produce, as output, reconstructed images that are stored as quantum states ("quantum images"). While reading out the individual pixels of the quantum image is not classically efficient, it may still be possible to extract useful information from quantum outputs if we use them as inputs to novel algorithms for post-processing reconstructed images, thereby maintaining the exponential speedup. 

Finally, though not discussed in detail here, quantum algorithms for image reconstruction open the path for quantum mechanical collection of medical imaging data. Since inputs to the quantum algorithms are wavefunctions, new experimental methods can be developed to collect or build this input wavefunction using fewer resources (\textit{e.g.}, time or radiation) and to input the resulting state directly into a quantum computer that can then perform the quantum image reconstruction algorithm.

\section{Acknowledgements}
We thank Milad Marvian, Giacomo de Palma, Lara Booth, Reevu Maity, Ryuji Takagi, and Zhaokai Li for helpful discussions and suggestions. This work was  supported by AFOSR, ARO, DOE, IARPA, and ARO.

\bibliography{main.bib} 

\pagebreak
\widetext
\begin{center}
\textbf{\large Supplemental Materials}
\end{center}
\setcounter{equation}{0}
\setcounter{figure}{0}
\setcounter{table}{0}
\setcounter{page}{1}
\makeatletter
\renewcommand{\theequation}{S\arabic{equation}}
\renewcommand{\thefigure}{S\arabic{figure}}
\renewcommand{\bibnumfmt}[1]{[S#1]}
\renewcommand{\citenumfont}[1]{#1}

\section{Classical Image Reconstruction Algorithms}
\label{classicalAlgs}
Common image reconstruction algorithms include: Fourier image reconstruction, filtered back-projection (FBP), and iterative reconstruction (IR) methods. 

Fourier image reconstruction makes use of the Fourier Slice Theorem, from which it follows that the image can be reconstructed by taking the inverse 2D Fourier transform of the 1D Fourier transformed Radon data. This can be mathematically summarized as \cite{aundalRadon1996}:

\begin{equation}
    \hat{F}(\nu \cos \theta, \nu \sin \theta) = \int_{-\infty}^{\infty}\hat{f}(r, \theta)e^{-2\pi i \rho \nu} d\rho
\end{equation}

\begin{equation}
    \binom{k_x}{k_y} = \nu \binom{\cos \theta}{\sin \theta}
\end{equation}

\begin{equation}
    f(x, y) = \int_{-\infty}^{\infty}\int_{-\infty}^{\infty}\hat{F}(k_x, k_y)e^{2\pi i(k_xx+k_yy)}dk_xdk_y
\end{equation}

\begin{figure}[H]
    \centering
    \includegraphics[scale = 0.8]{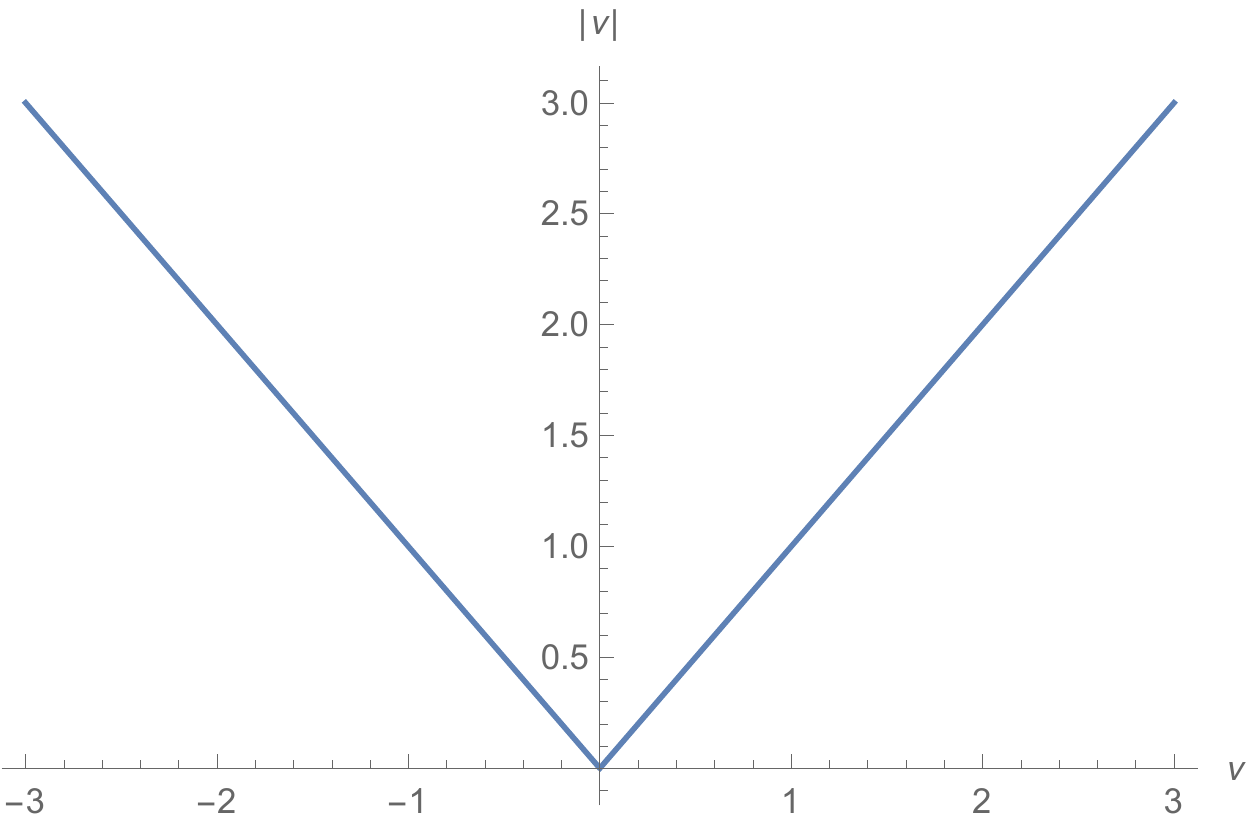}
    \caption{Ramp filter often used in filtered backprojection.} 
    \label{fig:rampFilter}
\end{figure}

In FBP, the radon data is first converted to the Fourier domain through a 1D Fourier transform. Because the data in the Fourier domain is highly concentrated around low frequencies, a high-pass filter, ideally the ramp-filter (figure \ref{fig:rampFilter}), is applied. An inverse 1D Fourier transform then recovers a filtered version of the original radon data. Applying a backprojection operator, $\mathcal{B}$, reconstructs the image \cite{aundalRadon1996}. 


The continuous definition of the Radon transform is \cite{pipatsrisawatPerformance2005}: 

\begin{equation}
    \hat{f}(r, \theta) = \int_{-\infty}^{\infty} \int_{-\infty}^{\infty} f(x, y)\delta(r-x\cos\theta - y \sin\theta)dx dy
\end{equation}

The Radon data is filtered in the Fourier domain \cite{aundalRadon1996}:

\begin{equation}
    \Tilde{f}(\rho, \theta) = \int_{-\infty}^\infty \lvert{\nu}\rvert\left(\int_{-\infty}^\infty \hat{f}(r, \theta)e^{-2\pi i r\nu}dr\right)e^{2\pi i \rho\nu}d\nu
\label{eq:rampFilter}
\end{equation}

The image is reconstructed by applying the backprojection operator, $\mathcal{B}$ \cite{aundalRadon1996, pipatsrisawatPerformance2005}:

\begin{equation}
    f(x,y) = \mathcal{B}\Tilde{f}(\rho, \theta)  = \int_0^\pi \Tilde{f}(x \cos{\theta} + y \sin{\theta}, \theta)d\theta = \int_0^\pi\int_{-\infty}^{\infty}\Tilde{f}(\rho, \theta)\delta(\rho - x\cos{\theta}-y\sin{\theta})d\rho d\theta
\end{equation}

Filtered back-projection alone, however, does not perform well on noisy data \cite{generalreconstruction2019}. IR algorithms rely on artificial data generated by forward projection (the reverse of back-projection) to provide correction terms for measured data \cite{generalreconstruction2019}. Many iterations allow for convergence towards corrected solutions, producing high quality images even on noisy data \cite{generalreconstruction2019}. Simultaneous Algebraic Reconstruction Technique (SART) is an IR algorithm that is frequently used in practice. It is a combination of two other iterative methods: Algebraic Reconstruction Technique (ART) and Simultaneous Iterative Reconstruction Technique (SIRT) \cite{sart1984}. ART methods apply a correction to each beam line integral individually, while SIRT methods use quadratic optimization to correct errors in all equations at once \cite{sart1984}. Although SIRT methods are less efficient than ART, they produce images that are less granulated than those resulting from ART-type methods \cite{sart1984}. SART aims to combine the benefits of both methods: to produce high quality images while minimizing the number of iterations necessary for convergence of the root-mean-squared error \cite{sart1984}. 






\section{Interpolation Methods}


Our interpolation step is a grid transformation from polar to Cartesian within the frequency domain. Some common interpolation methods are: bilinear, cubic B-spline, nearest neighbor, and simplex. For our purposes, we are restricted to linear techniques due to unitary constraints. Bilinear interpolation (which can be scaled up to interpolate from \textit{k} nearest points) is our method of choice. Bilinear interpolation, as compared to nearest neighbor and simplex interpolation, maintains more information about the polar neighborhood of a Cartesian pixel, and is therefore favorable for our image reconstruction method. Nearest neighbor, simplex, and bilinear interpolation are defined below for interpolations of $f(x', y')$ to $f(x,y)$.


\begin{enumerate}

\item In nearest neighbor interpolation, some pixel $p = f(x', y')$ is assigned a new value by taking the value of the closest input sample (rounding to the nearest integer) \cite{interpolation2011}:

\begin{equation}
    f(x, y) = f(\lfloor x' \rceil, \lfloor y' \rceil)
\end{equation}

\item Simplex interpolation assigns interpolation weights using triangles generalized to n dimensions. In the case of 2D image reconstruction methods, the problem is simplified to 2-simplex interpolation, which uses regular triangles as shown in figure \ref{fig:simplex}. A point $P = (x, y)$ with value $f(x, y)$ lying within a triangle \textit{ABC} with vertex values $f(x_A, y_A), f(x_B, y_B), f(x_C, y_C)$ forms three new triangles: $ABP$, $ACP$, $BCP$. The weight of each vertex value is defined by the ratio of each subtriangle area to the total area \cite{VirtanenScipy2020}:

\begin{equation}
    f(x, y) = \frac{area(ABP)f(x'_C, y'_C) + area(ACP)f(x'_B, y'_B) + area(BCP)f(x'_A, y'_A)}{area(ABC)}
\end{equation}

\begin{figure}[ht]
    \centering
    \includegraphics[scale = 0.4]{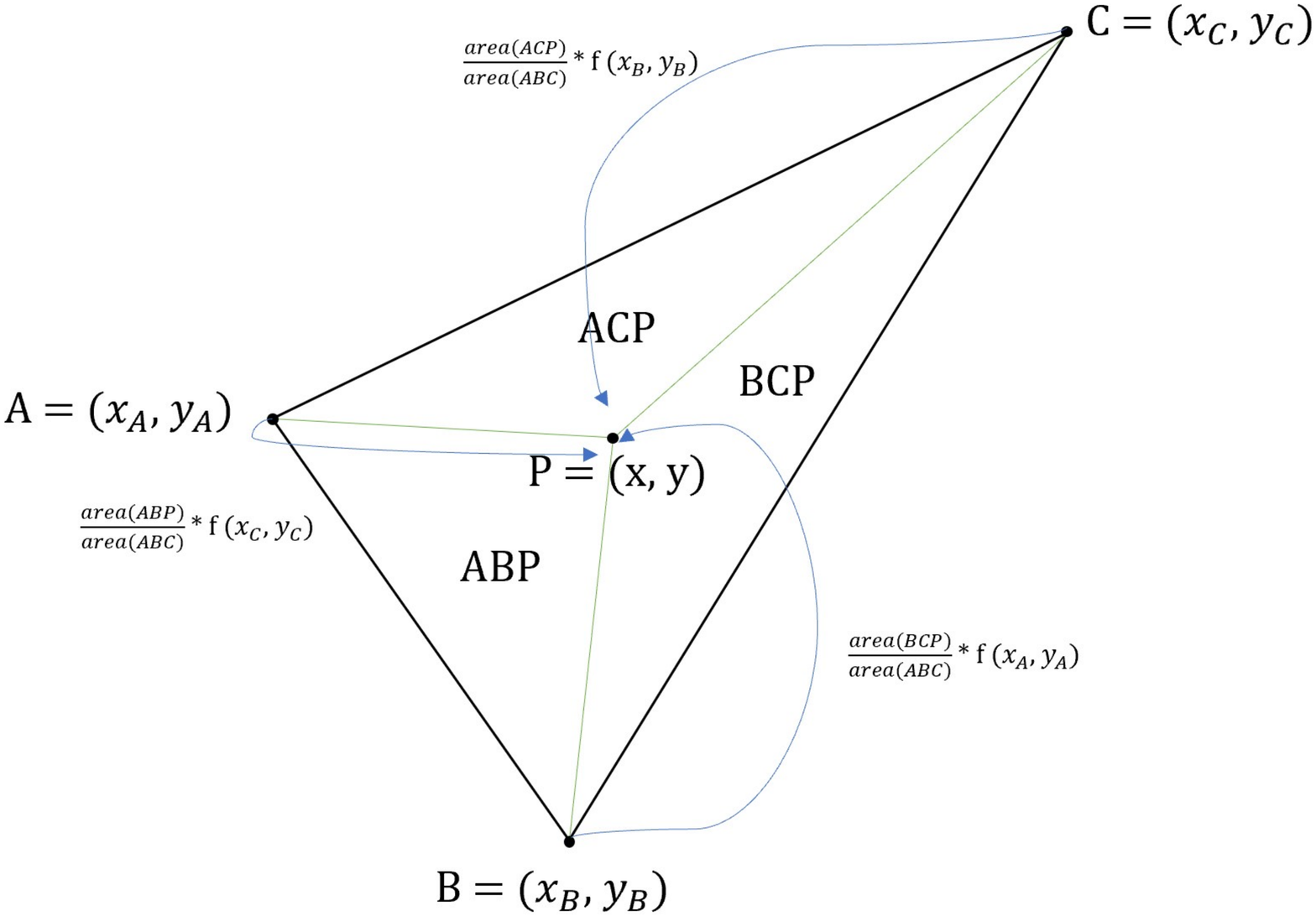}
    \caption{Schematic representation of 2-simplex interpolation. Interpolation value at point P is assigned based on the areas of triangles containing P as a vertex.}
    \label{fig:simplex}
\end{figure}


\item Bilinear interpolation is a linear interpolation in two directions. First, the four nearest neighbors are found for some point $(x', y')$ and weights $w_x, w_y$ are calculated:

\begin{equation}
    a = \floor*{\frac{x'}{\Delta x}} \rightarrow x_a \leq x' \leq x_{a+1}, w_x = \frac{x' - x_a}{\Delta x}
\end{equation}
\begin{equation}
    b = \floor*{\frac{y'}{\Delta y}} \rightarrow y_b \leq y' \leq y_{b+1}, w_y = \frac{y' - y_b}{\Delta y}
\end{equation}

These are then substituted into the interpolation equation \cite{aundalRadon1996}:

\begin{equation}
    f(x, y) = (1 - w_y)((1-w_x)f(x_a, y_b) + w_xf(x_{a+1}, y_b)) + w_y((1-w_x)f(x_b, y_{b+1})+w_xf(x_{a+1}, y_{b+1}))
\end{equation}

\end{enumerate}




\section{Fourier Slice Reconstruction in Three Dimensions}

Extending the Fourier Slice Theorem for reconstruction in three dimensions is straightforward and provided briefly here. For a more complete analysis and discussion of reconstruction via the Fourier Slice Theorem in three dimensions, we recommend chapter 10 in \cite{aundalRadon1996}. In fact, the brief overview provided here follows closely with \cite{aundalRadon1996}.

For our purposes, we use the following parameterization to specify a line $\boldsymbol{r}$ in 3D. Vectors are indicated in bold font.

\begin{equation}
    \boldsymbol{r} = \boldsymbol{r} + s \boldsymbol{\tau}
\end{equation}
where $\boldsymbol{r}$ is an offset vector and $s$ is a scalar that spans over the given line. $\boldsymbol{\tau}$ is a unit vector parameterized by angles $\theta$ and $\phi$ as is common in specifying points on a unit sphere.

\begin{align}
    \boldsymbol{\tau} &= \begin{bmatrix}
          \cos{\theta}\cos{\phi} \\
          \sin{\theta}\cos{\phi} \\
          \sin{\phi}
         \end{bmatrix}
\end{align}

To make things simple, we assume that $\boldsymbol{r_0}$ is spanned by two basis vectors $\boldsymbol{\alpha}$ and $\boldsymbol{\beta}$ orthogonal to $\boldsymbol{\tau}$. 

\begin{equation}
    \boldsymbol{r_0} = u \boldsymbol{\alpha} + v \boldsymbol{\beta}
\end{equation}

In these coordinates, the Radon transform can be written as follows.

\begin{dmath}
    \mathcal{R}F(\boldsymbol{r}) = f(\theta, \phi, u, v) = \int_{-\infty}^{\infty} F( s' \boldsymbol{\tau} + u \boldsymbol{\alpha} + v \boldsymbol{\beta} )  \,ds'
\end{dmath}

With this notation, we can now provide the Fourier slice theorem in three dimensions.

\begin{theorem}[Fourier slice theorem in three dimensions \cite{aundalRadon1996}]
\label{slice3d}
Let $\hat{F}(\boldsymbol{k})$ be the three dimensional Fourier transform of  $F(\boldsymbol{r})$ and $\hat{f}(k_\alpha \boldsymbol{\alpha} + k_\beta \boldsymbol{\beta}, \theta, \phi)$ be the two dimensional Fourier transform of $f(\theta, \phi, u \boldsymbol{\alpha} + v \boldsymbol{\beta})$ over the $\boldsymbol{\alpha}$ and $\boldsymbol{\beta}$ dimensions. The values of $f(\theta', \phi', u \boldsymbol{\alpha} + v \boldsymbol{\beta})$ form a plane passing through the origin perpendicular to the line given by $\theta'$ and $\phi'$ and are equal to $\hat{F}(\boldsymbol{k})$ in the same plane. 
\begin{equation}
    F(\boldsymbol{r}) = \int_{-\infty}^{\infty} \int_{-\infty}^{\infty} \int_{-\infty}^{\infty} \hat{F}(\boldsymbol{k}) e^{-i2\pi \boldsymbol{r \dot k}} \,d \boldsymbol{k}
\end{equation}
\begin{equation}
    \hat{F}(\boldsymbol{k}) = \int_{-\infty}^{\infty} \int_{-\infty}^{\infty} f(\theta, \phi, u \boldsymbol{\alpha} + v \boldsymbol{\beta}) e^{-i2\pi(u k_\alpha + v k_\beta)} \,du\,dv
\end{equation}
\end{theorem}

Given the above, reconstruction in 3D follows the same order of steps as in algorithm \ref{alg:classical} with proper adjustment. For step \ref{alg:classical1dfft}, 1D Fourier transform over lines are replaced by 2D Fourier transforms over planes. Interpolation (step \ref{alg:classicalInterpolation}) is now performed in 3 dimensions. Finally, to recover the original image (step \ref{alg:classical2dfft}), a 3D inverse Fourier transform replaces the inverse 2D Fourier transforms.

\section{Classical Implementation of Algorithm 1}


We have implemented a classical simulation (Algorithm 1) of our quantum algorithm using Python. The reconstructed image is reported in figure S1. The source code will be available upon publication. 

Steps 1 and 3 of Algorithm 1 were completed using built-in methods from SciPy \cite{scipy2014}.
The novelty of our implementation is the interpolation, Step 2. This was done using sparse matrix bilinear interpolation, since sparsity is a requirement for the quantum version of our algorithm. Algorithm 3 describes this step in detail.
\begin{algorithm}[H]
\caption{Sparse Matrix Bilinear Interpolation}\label{alg:bilinear}
\textbf{Input: } 
\begin{itemize}
    \item Edge length of output image, in pixels: \textit{S}
    \item A one dimensional array of pixel values at each pixel location, in polar coordinates: \textit{vals}  
\end{itemize} 
\textbf{Result: } Sparse matrix containing interpolated pixel values, where each pixel location is in Cartesian coordinates: \textit{res}

\begin{algorithmic}[1]
\State Initialize empty sparse matrix with appropriate dimensions
\Comment{returns \textit{\textit{M}}}
\State Initialize an array of Cartesian points in appropriate order 
\Comment{returns \textit{CartesianGrid}}
\State \textbf{for} point in \textit{CartesianGrid}:
\State \quad get interpolation weights 
\Comment {see description below}
\State \quad assign each weight as an entry in the row of M that corresponds to the current Cartesian point
\State \textbf{return} \textit{M} matrix multiplied with the vector \textit{vals}
\end{algorithmic}
\end{algorithm}

Step 4 in Algorithm 3 requires some additional details. First, the specified Cartesian point is converted to polar coordinates: $(k_p, \theta)$. It is important to note that because projected data is only between values of 0 and $\pi$, this conversion step must restrict each Cartesian point to stay within this range. 

Next, the nearest upper and lower integer values are found for $(k_p, \theta)$: $k_{p_h}, k_{p_l}, \theta_h, \theta_l$, leading to 4 points: $(k_{p_h}, \theta_h), (k_{p_l}, \theta_l), (k_{p_h}, \theta_l), (k_{p_l}, \theta_h)$. The weights for each point are assigned as follows \cite{aundalRadon1996}:

$$w_\theta = \frac{\theta-\theta_l}{\Delta \theta}, \quad w_k = \frac{k-k_l}{\Delta k}
$$
$$
\begin{matrix}
(k_{p_l}, \theta_l) \rightarrow (1-w_\theta)(1-w_k)\\
(k_{p_h}, \theta_l) \rightarrow (1-w_\theta)(w_k)\\
(k_{p_l}, \theta_h) \rightarrow (w_\theta)(1-w_k)\\
(k_{p_h}, \theta_h) \rightarrow (w_\theta)(w_k)
\end{matrix}
$$

These four points correspond to four entries in a row of \textit{M}, and are incorporated into the algorithm as described in Step 5 of Algorithm 3.

\begin{figure}[ht]
    \centering
    \includegraphics[scale = 0.5]{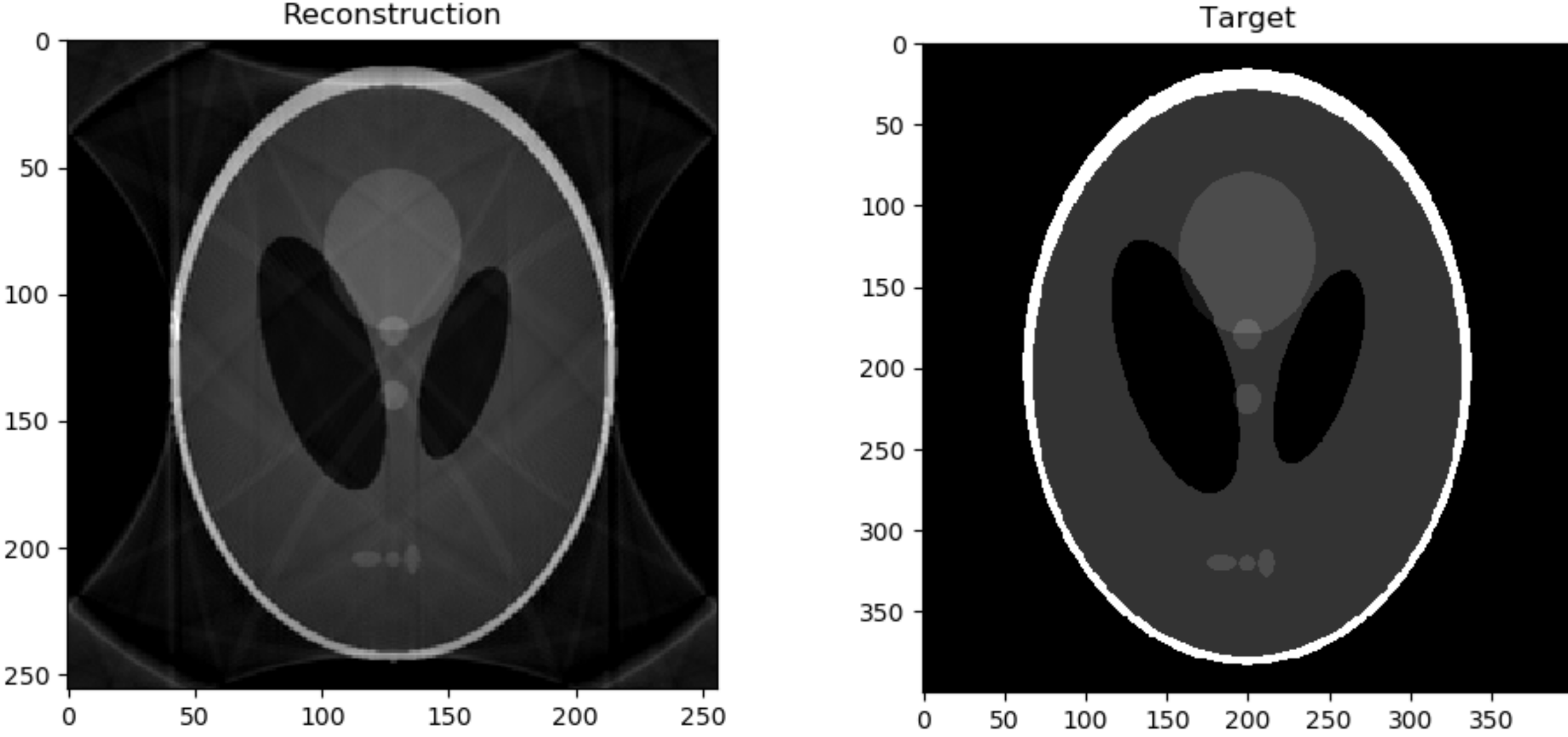}
    \caption{Reconstructed Shepp Logan Phantom using our classical implementation (left) and original Shepp Logan Phantom (right).}
    \label{fig:resultsComp}
\end{figure}

Figure \ref{fig:resultsComp} shows the reconstruction results from our implementation \cite{Shepploganhead1974}. It is important to note the artifacts: the streaks likely result from interpolation within the Fourier domain. One potential cause could be lack of accurate high frequency data. High frequency information within the Fourier domain leads to fine details in the spatial domain. Because of the structure of the projected data, there is a maximum frequency for which pixel values exist. When interpolating, however, the Cartesian grid extends beyond these values, and thus it is difficult to get accurate high frequency measurements. This feature is why Fourier interpolation is often not used in classical image reconstruction algorithms. We use Fourier interpolation for quantum image reconstruction because the quantum Fourier transform gives an exponential speedup over the classical Fourier transform.

\section{Quantum Interpolation}
The interpolation step converts the data in the frequency domain from polar to Cartesian coordinates. In polar coordinates, the data $\hat{f}(k_\rho,\theta)$ is given at discrete, equally spaced values of $k_\rho$ and $\theta$. We desire the data in Cartesian coordinates $\hat{F}(k_x,k_y)$ for discrete, equally spaced values of $k_x$ and $k_y$. We perform the following operation.

\begin{dmath}
    e^{-iA^* } \ket{1}\hat{f}(k_\rho,\theta)\ket{k_\rho}\ket{\theta} = \cos{(A^* t)} \ket{1}\hat{f}(k_\rho,\theta)\ket{k_\rho}\ket{\theta} + i \sin{(A^* t)} \ket{1}\hat{f}(k_\rho,\theta)\ket{k_\rho}\ket{\theta} 
    \approx \ket{1}\hat{f}(k_\rho,\theta)\ket{k_\rho}\ket{\theta} + it \ket{0}\hat{F}(k_x,k_y) \ket{k_x}\ket{k_y} 
\end{dmath}

Here, we recall that we constructed the Hermitian matrix $A^*$ to perform the interpolation, since a given linear interpolation matrix $A$ is not necessarily Hermitian:
\begin{equation}
    A^* = \sigma^- \tens{} A + \sigma^+ \tens{} A^\dagger = 
    \begin{pmatrix}
    0 & A\\
    A^\dagger & 0
    \end{pmatrix}
\end{equation}

\begin{dmath}
    A^* \ket{1}\hat{f}(k_\rho,\theta)\ket{k_\rho}\ket{\theta} = 
    \begin{pmatrix}
    0 & A\\
    A^\dagger & 0
    \end{pmatrix}
    \begin{pmatrix}
    0\\
    \hat{f}(k_\rho,\theta)
    \end{pmatrix}
    =
    \begin{pmatrix}
    \hat{F}(k_x,k_y) \\
    0
    \end{pmatrix}
\end{dmath}

We choose $t$ to output a state that is $\epsilon$ close to $A^* t$ when measuring $\ket{0}$ on the ancillary qubit.

\begin{dmath}
    \Vert \sin{(A^*t)} - A^* t \Vert \leq \epsilon
\end{dmath}

We can bound the above by truncating the Taylor series:
\begin{dmath}
    \Vert \sin{(A^*t)} - A^* t \Vert \leq \frac{1}{6} \Vert A^* \Vert ^3 |t|^3
\label{eq:errorTruncation}
\end{dmath}

Since $A$ is an interpolation matrix, we now proceed to bound the singular values of $A^*$ by showing that the number of points within an interpolation region is bounded. Given an $m \times m$ matrix $A$ with entries $a_{i j}$, we can bound the singular values $\sigma$ using Schur's bound\cite{nikiforovRevisiting2007}:

\begin{dmath}
    \sigma^2(A) \leq \max_{i \in [m], j \in [m]} r_i c_j,
\end{dmath}
where $r_i = \sum_{k \in [m]} |a_{i k}|$ and $c_j = \sum_{k \in [m]} |a_{k j}|$.

Each row of $A$ interpolates from $s$ different polar coordinates into one Cartesian coordinate. The sum of the entries of any row in the matrix $A$ equals 1 ($r_i = 1, \forall i$). The column sum $c_j$ can be bounded by considering the maximum number of points in the Cartesian grid which require a specific discretized point in the polar grid to perform interpolation. The polar grid has largest influence on the outer boundaries where its radius is largest. To calculate the number of Cartesian points that are within the influence of a polar point on the outside of the polar grid, we consider the case of bilinear interpolation. The Cartesian grid has $N$ equally spaced grid points in each dimension with $\Delta x = \Delta y = 1/N$. Similarly, the polar grid has $N$ equally spaced points in the $\rho$ and $\theta$ dimensions with $\Delta \rho = \pi / N$ and $\Delta r = 1/N$. Here, a Cartesian point $(x',y')$ is influenced by a specified polar point $(\rho ', \theta ')$ if $\rho ' - \Delta \rho \leq \sqrt{x^{'2} + y^{'2}} \leq \rho ' + \Delta \rho$ and $\theta ' - \Delta \theta \leq \arctan{|\frac{y'}{x'}|} \leq \theta ' + \Delta \theta$. At the limit where $N$ is large, this region approaches the shape of a rectangle with edges of length $2 / N$ and $2 \pi / N$. These edge lengths can fit at most $3$ and $7$ points spaced $1/N$ apart respectively. Thus, a maximum of $21$ equally spaced Cartesian points can fall within this region  ($c_i \leq 21, \forall i$). Combining the above results, we note that the singular values are bounded:

\begin{dmath}
    \sigma^2(A) \leq 21,
\label{eq:singValMax}
\end{dmath}

Combining equations \ref{eq:errorTruncation} and \ref{eq:singValMax}, we find that it is possible to choose a $t$ for any $N$ to limit the error in our output state to $\epsilon$. 

Finally, for small $\epsilon$, the probability of successfully measuring the ancillary qubit in the state $\ket{0}$ is equal to $p_0 = \bra{k_y}\bra{k_x}\hat{F}(k_x,k_y)^\dagger \hat{F}(k_x,k_y) \ket{k_x}\ket{k_y}$. This probability will depend on the nature of the problem and how the interpolation is performed. In all cases of sparse interpolation, low frequency elements are partially in the kernel of the interpolation matrix since there are more low frequency data points in the polar representation compared to the Cartesian representation. In fact, since reconstruction of a continuous signal or image by the Fourier slice theorem is equivalent to reconstruction by filtered back-projection, any ideal interpolation method would also perform a ramp filter on the frequency components (see equation \ref{eq:rampFilter} and figure \ref{fig:rampFilter}). Given this filter, it is clear that the interpolation step is efficient in cases where the data is not dominated by the ill conditioned subspace (low frequencies) that needs to be filtered out. Thus, if the portion of the data within the low frequency components does not grow with the size of an image, the implementation proposed here scales efficiently with the size of an image.

\end{document}